\documentclass[aps,prbrc,twocolumn,amsmath,amssymb,showpacs, superscriptaddress,notitlepage,longbibliography]{revtex4-1}

\usepackage[colorlinks=true,linkcolor=blue,anchorcolor=red,citecolor=blue, urlcolor=blue]{hyperref}
\usepackage{bm}
\usepackage{graphicx}
\usepackage{color}
\usepackage{cases}
\usepackage{bbm}
\usepackage{MnSymbol}
\usepackage{pifont}

\begin{document}

\title{Quantum oscillation beyond the quantum limit in pseudospin Dirac materials}

\author{C. M. Wang}

\affiliation{Department of Physics, Shanghai Normal University, Shanghai 200234, China}
\affiliation{Institute for Quantum Science and Engineering and Department of Physics, Southern University of Science and Technology, Shenzhen 518055, China}
\affiliation{Shenzhen Key Laboratory of Quantum Science and Engineering, Shenzhen 518055, China}

\author{Hai-Zhou Lu}
\email{Corresponding author: luhz@sustech.edu.cn}
\affiliation{Institute for Quantum Science and Engineering and Department of Physics, Southern University of Science and Technology, Shenzhen 518055, China}
\affiliation{Shenzhen Key Laboratory of Quantum Science and Engineering, Shenzhen 518055, China}

\author{X. C. Xie}
\affiliation{International Center for Quantum Materials, School of Physics, Peking University, Beijing 100871, China}
\affiliation{CAS Center for Excellence in Topological Quantum Computation, University of Chinese Academy of Sciences, Beijing 100190, China}
\affiliation{Beijing Academy of Quantum Information Sciences, West Building 3, No. 10, Xibeiwang East Road, Haidian District, Beijing 100193, China}

\begin{abstract}
Recently, many unexpected fine structures in electric, magnetic, and thermoelectric responses at extremely magnetic fields in topological materials have attracted tremendous interest.
We propose a new mechanism of quantum oscillation beyond the strong-field quantum limit for Dirac fermions. The amplitude of the oscillation is far larger than the usual Shubnikov--de Haas oscillation. The oscillation tends to be periodic in the magnetic field $B$, instead of $1/B$. The period of the oscillation does not depend on the Fermi energy. These behaviors cannot be described by the famous Lifshitz-Kosevich formula. The oscillation arises from a mechanism that we refer to as the inversion of the lowest Landau level, resulted from the competition between the pseudospin Dirac-type Landau levels and real-spin Zeeman spitting beyond the quantum limit. This inversion gives rise to the oscillation of the Fermi energy and conductivity at extremely large magnetic fields. This mechanism will be useful for understanding the unexpected fine structures observed in the strong-field quantum limit in Dirac materials.
\end{abstract}

\date{\today}

\maketitle

{\color{blue}\emph{Introduction}} --
In a strong magnetic field, the energy spectrum of non-interacting electrons is quantized into the Landau levels. The spacing and degeneracy of the Landau levels increase with increasing magnetic field, so the Fermi energy (which marks the highest energy the electrons occupy at) crosses the Landau levels one by one, leading to the quantum oscillations of the resistivity (Shubnikov-de Haas) and magnetization (de Haas-van Alphen). In an extremely strong magnetic field the system can enter the quantum limit (QL) in which the carriers occupy only the lowest zeroth Landau level. In the QL, the quantum oscillation is supposed to cease because there is no lower Landau level that allows the Fermi energy to move onto. In the newly discovered 3D topological semimetal \cite{Wan11prb,Yang11prb,Burkov11prl,Xu11prl,Delplace12epl,Jiang12pra,Young12prl,Wang12prb,Singh12prb,Wang13prb,LiuJP14prb,Bulmash14prb}, 
unexpected oscillation-like fine structures have been reported in the QL and cannot be understood by the theory of quantum oscillation, such as magnetic-tunnelling-induced Weyl node annihilation \cite{ZhangCL17np, Ramshaw18nc}, breakdown of the chiral anomaly \cite{Kim2017prl}, log-periodic oscillations \cite{wang2018sa}, forbidden backscattering \cite{ChenYY18prl,Assaf17prl}, anomalous thermoelectric effects \cite{Zhang2019prl,Kozii2019prb,zhang2020nc,han2019arXiv}, and the magnetic torque and magnetization anomalies \cite{Moll16nc, ZhangCL19nc}.

In this Rapid Communication, we propose a theory of quantum oscillation beyond the QL, due to a mechanism that we refer to as the inversion of the lowest Landau level (ILLL), i.e., the lowest Landau band may exchange
their positions in energy at a critical magnetic field in topological Dirac materials where only pseudospin is coupled to momentum. Because usually oscillations are not supposed to happen at the quantum-limit magnetic fields, we refer to them as the quantum oscillations beyond the QL.
This inversion does not need the topological band inversion in the absence of magnetic field \cite{Konig07sci,ZhangSB15srep}.
This ILLL induced oscillation beyond the QL is absent for Schr\"odinger electrons even in the presence of Zeeman interaction or spin-orbital coupling \cite{Bychkov1984jpc,Wang03prb}.  
The new quantum oscillation can be measured in the conductivity.
This energy spectrum driven mechanism is distinct from the usual Shubnikov-de Haas oscillation (SdHO) in several aspects (Table \ref{Tab:Comparison}). The oscillation due to the ILLL beyond the QL is periodic in the magnetic field $B$, not in $1/B$ as the SdHO is. The frequency of the oscillation does not depend on the Fermi energy. We use a 2D massless graphene-like Dirac model to illustrate the ILLL beyond the QL and related observables. 
We further generalize it to 3D Dirac materials. It will be helpful to understand the unexpected fine structures in conductivity, magnetization, and thermoelectric properties in extremely strong magnetic fields.

\begin{table}[htbp]
\caption{The comparison between the oscillation due to the ILLL beyond the QL and the usual SdHO. $B$ is the magnetic field. $E_F$ is the Fermi energy. }
\begin{ruledtabular}
\begin{tabular}{ccc}
 & Shubnikov-de Haas & ILLL  \\
\hline
Quantum limit & Before  & Beyond
\\
$B$-dependence & Periodic in $1/B$  & Periodic in $B$
\\
Frequency  & depends on $E_F$   & does not depend on $E_F$
\tabularnewline
\end{tabular}\label{Tab:Comparison}
\end{ruledtabular}
\end{table}

\begin{figure}[htbp]
\centering
\includegraphics[width=0.92\columnwidth]{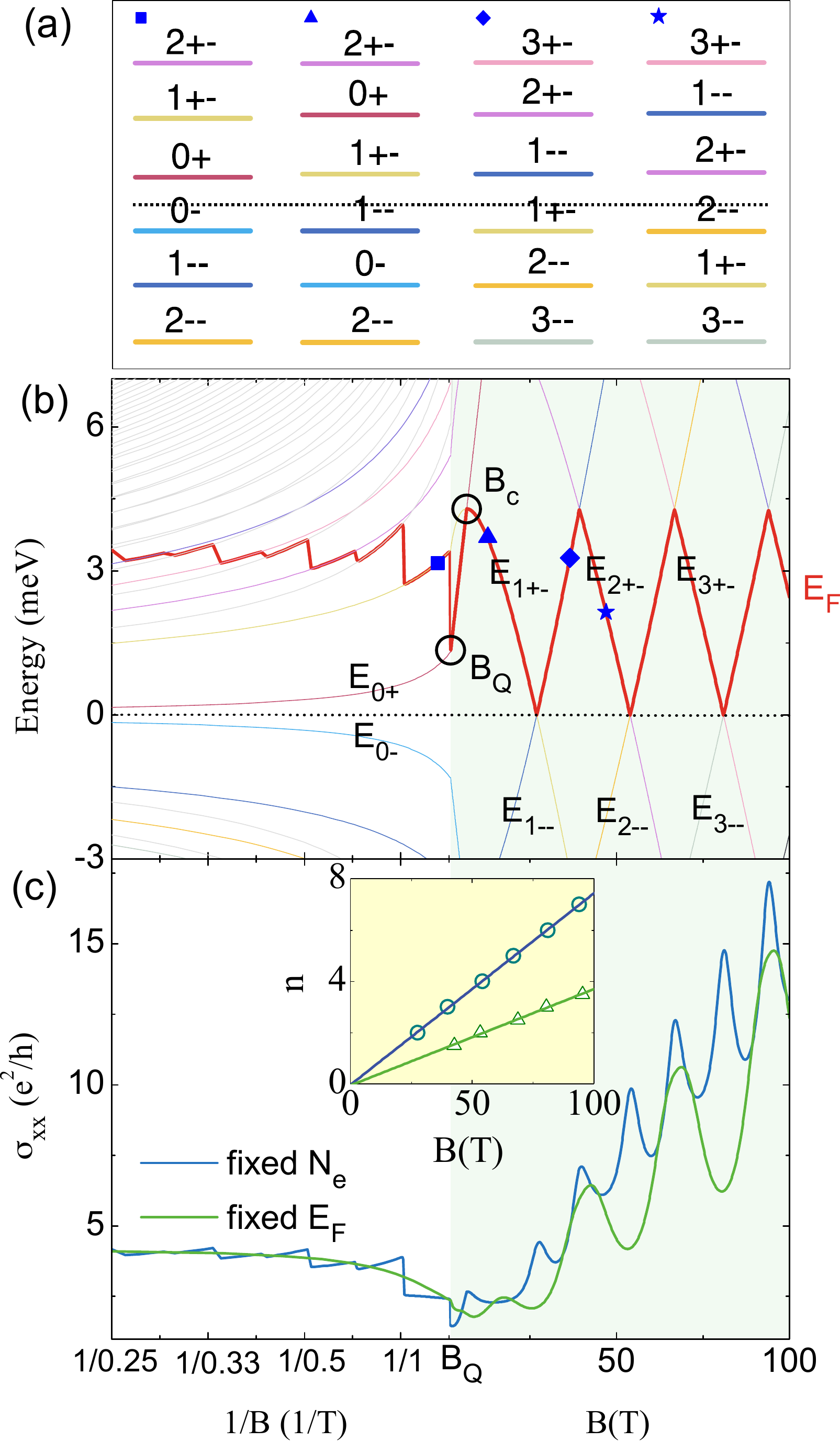}
\caption{(Color online) The quantum oscillations beyond the QL (green shadow) induced by the ILLL for the 2D graphene-like model in Eq. (\ref{Eq:Model2D}). (a) The energies of the Landau levels at magnetic fields marked by $\filledmedsquare$, $\filledmedtriangleup$, \ding{117}, and \ding{72} in (b). The dash line indicates the location of the zero energy. (b) The Landau levels and Fermi energy $E_F$ (red) as functions of $B$ for a fixed electron density $N_e$. $\bigcirc$ mark the critical fields where $E_F$ is moving to $E_{0+}$ ($B_Q$, the QL) and from $E_{0+}$ to $E_{1+-}$ ($B_c$), respectively. (c) The conductivity $\sigma_{xx}$ as a function of the magnetic field for a fixed $N_e$ (blue) or a fixed zero-field Fermi energy $E_F^0$ (green). Before entering the QL (left side of $B_Q$), there is the usual SdHO. The inset shows the peaks or dips positions versus the Landau index beyond the QL for fixed density (circle) and Fermi energy (triangle). The parameters are $v=0.6$ eV\AA, $g=22$, $N_e=5\times10^{10}$ cm$^{-2}$, and the broadening of levels $\Gamma=0.002$ eV.
}\label{Fig:2D}
\end{figure}

{\color{blue}\emph{Inversion of the lowest Landau level beyond the QL -- 2D graphene-like model}} --
In order to illustrate the ILLL, we consider a $4\times 4 $ graphene-type model in a $z$-direction magnetic field $B$
\begin{eqnarray}\label{Eq:Model2D}
H_{\rm 2D}= v[(k_x+\frac{yeB}{\hbar})\tau_x+k_y\tau_y]\otimes\mathbbm{1}_2+\frac{g\mu_BB}{2} \mathbbm{1}_2\otimes\sigma_z,\nonumber\\
\end{eqnarray}
where $v$ is the velocity, $k_x,k_y$ are the wave vectors, $y$ is the position operator, the Pauli matrices $(\tau_x,\tau_y)$ and $\sigma_z$
stand for the $x,y$-direction pseudospin and $z$-direction real spin, respectively, $\mathbbm{1}_2$ is the $2\times2$ unit matrix, and the last term is the Zeeman splitting with the effective $g$ factor and Bohr magneton $\mu_B$. The cyclotron motion driven by $B$ quantizes the energy spectrum of the pseudospin part into the Landau levels, while the real-spin part takes the Zeeman splitting. The energy spectrum can be found as 
\begin{eqnarray}\label{Eq:E_nu}
E_{\nu \lambda s}=\lambda\left(v\sqrt{\frac{2\nu e}{\hbar}}  \sqrt{B}+s \frac{g\mu_B}{2} B\right),
\end{eqnarray}
for Landau indices $\nu=1,2,3,\cdots$ and $\lambda,s=\pm1$; and
\begin{eqnarray}
E_{0\pm}=\pm\frac{g\mu_B}{2} B,
\end{eqnarray}
for $\nu=0$.
Because of the Dirac nature of the pseudospin part, the energies of the Landau levels scale with
$\sqrt{B}$ \cite{Neto09rmp}. The Zeeman splitting, on the other hand, is linear in $B$. Figure  \ref{Fig:2D} (a) shows the energies of the Landau levels as functions of the magnetic field. There are two distinct features compared to usual cases. First, beyond a critical magnetic field (marked as $B_c$), the $\nu=0$ Landau levels $E_{0\pm}$ are no longer the lowest in energy (closest to zero energy). Instead, $E_{1\lambda -}$ become the lowest, then $E_{2\lambda -}$, $E_{3\lambda -}$, ... become the lowest one by one with increasing magnetic field. We refer to this as the inversion of the lowest Landau level, which arises because the linear-$B$ Zeeman splitting overwhelms the $\sqrt{B}$ Landau-level energies. The critical magnetic field required to inverse $E_{0\lambda }$ and $E_{1\lambda -}$ can be found as
\begin{eqnarray}\label{Eq:Bc}
B_c &=& \frac{e}{\hbar} \frac{2v^2}{g^2\mu_B^2}.
\end{eqnarray}
Above this field is what we referred to as ``beyond the quantum limit". This ILLL is not observed in graphene, in which $g=2$ and $v=1.1\times 10^6$ m/s (7 eV\AA) so $B_c$ is about 1.2 $\times$ 10$^5$ T. If $g=20$, $v=1.1\times 10^5$ m/s (0.7 eV\AA), this critical field can be greatly reduced to 12 T, which is accessible to high-magnetic field laboratories. The second feature is that for $\nu\ge1$ the energies $E_{\nu \lambda -}$ cross when the first term is equal to the second term in Eq. (\ref{Eq:E_nu}).

{\color{blue}\emph{Fermi energy oscillation beyond the QL}} --
For samples with a large density, the fixed electron density case could be used in most solids. Since the solid is always charge neutral and the current flowing into the sample is in balance with the one flowing out it, the total electron density is a constant \cite{Mahan1990}. If the samples with small number of electrons contact to the metallic reservoirs, the most likely fixed one is the Fermi energy. For the former case, the Fermi energy $E_F$ needs to be determined by how a fixed electron density $N_e$ occupies the Landau levels \cite{ZhangCL19nc}, according to
\begin{align}
N_e=\frac{eB}{2\pi \hbar }\left[ n_{\rm F}(E_{0+})+\sum_{\nu,\lambda,s}^{E_{\nu \lambda s}>0}n_{\rm F}(E_{\nu \lambda s})\right].
\end{align}
Here $n_{\rm F}(x)=[\exp((x-E_F)/k_B T)+1]^{-1}$ is the Fermi distribution function. Figure \ref{Fig:2D} (b) shows the calculated Fermi energy as a function of the magnetic field for a fixed $N_e$.
Before entering the QL (the white regime), the Fermi energy is almost unchanged with increasing magnetic field. For usual cases, the Landau levels remain in the same order, the Fermi energy will be pushed down onto the Landau level with the lowest index (conventionally denoted as $0$th) and stays there at higher magnetic fields, which is the QL.
In contrast, here the ILLL and periodic crossing of $E_{\nu \lambda -}$ lead to the strong oscillation of the Fermi energy beyond the QL. $E_F$ shows dips where the energies $E_{\nu \lambda -}$ ($\nu\ge1$) cross. The period of the oscillation of $E_F$ is found as
$B_T = 4B_c$.

{\color{blue}\emph{Conductivity oscillation beyond the QL}} --
This oscillation beyond the QL can be measured in the conductivity. We calculate the longitudinal conductivity $\sigma_{xx}$ with the help of the Kubo formula \cite{Supp-Weyl}
$\sigma_{xx}=-(\hbar e^2 /2\pi V)\int d\varepsilon (\partial n_{\rm F} /\partial \varepsilon)
\text{Tr} (v_x  G^R  v_x  G^A  )$,
where $V$ is the area for 2D or the volume for 3D system, $G^{R/A}$ are the retarded and advanced Green's functions, respectively, and $v_x$ is the $x$-direction velocity operator.

Figure \ref{Fig:2D} (c) shows the longitudinal conductivities for fixed electron density $N_e$ and zero-field Fermi energy $E_F^0$, respectively. Before entering the QL, there is the usual SdHO of the conductivity (the white regime), periodic in $1/B$.
More importantly, a much stronger oscillation appears beyond the QL (the green shadow) for both two cases. The oscillation beyond the QL is periodic in the magnetic field $B$, not in $1/B$ as the usual SdHO is. The period for fixed $E_F^0$ is $B_T=4B_c $. However, for fixed density, the frequency of conductivity doubles compared with that of the Fermi energy, because the Dirac fermion is massless. If the Dirac fermion has no mass, the Landau levels $E_{\nu \pm -}$ can cross, so the Fermi energy can drop to zero and cross the negative level $E_{\nu - -}$ (see Fig. \ref{Fig:2D} (b)). As a result, conductivity peaks also occur at the zeros of the Fermi energy. The period of the conductivity beyond the QL can thus be found as $B_T=2B_c $.  The period of the oscillation beyond the QL does not depend on the Fermi energy, quite different from the usual SdHO \cite{WangCM16prl,LiC18prl}. It only depends on the ratio of the Fermi velocity and the $g$ factor, which determines the ILLL. This novel linear-$B$ oscillation gives a zero phase shift in the $n-B$ Landau fan diagram if one assigns integer numbers to the peaks or dips of conductivity, as shown in the inset of Fig. \ref{Fig:2D} (c). This is valid for both fixed Fermi energy and electron density. Previously, linear-$B$ oscillations are usually due to the quantum coherent transport such as Aharonov-Bohm effect or Altshuler-Aronov-Spivak effect. Here we demonstrate a new linear-$B$ oscillation.

All these feature could be understood from the analytical results of the density of states before and beyond the QL written as \cite{Supp-Weyl}
\begin{align}
  \varrho(E_F\gg E_{0+})
  \simeq&\frac{E_F}{\pi v^2}+A_1\cos\left(2\pi\frac{\hbar E_F^2}{2v^2eB}\right),\\
  \varrho(E_F\ll E_{0+})
  \simeq&\frac{g\mu_BB}{2\pi v^2}+A_2\cos\left(2\pi\frac{\hbar g^2\mu_B^2B}{8ev^2}\right).
\end{align}
Here amplitudes $A_1=\frac{2E_F}{\pi v^2}\exp\left(-\frac{2\pi \hbar E_F\Gamma}{v^2eB}\right)$ and $A_2=\frac{g\mu_BB}{\pi v^2}\exp\left(-\frac{\pi\hbar g\mu_B\Gamma}{ev^2}\right)$. For $E_F\gg E_{0+}$ (before the QL), the density of states oscillates inversely with the field as the usual SdHO. However, beyond the QL ($E_F\ll E_{0+}$), the density of states oscillates with the magnetic field having a period $B_T=4B_c$. Further, the phase shift of this $B$-oscillation equals zero for both two cases. The whole density of states beyond the QL $\varrho(E_F\ll E_{0+})$ is independent of the Fermi energy. The amplitude of the usual oscillation before the QL  increases with the field due to the increase of the Dingle factor. However, the amplitude beyond the QL is even larger since the ratio 
\begin{align}
\frac{A_2}{A_1}=\frac{E_{0+}}{E_F}e^{-2\pi\hbar(E_{0+}-E_F)\Gamma/(v^2eB)}>1,
\end{align}
because of the factor $E_{0+}\gg E_F$ and small broadening $\Gamma$, $e^{-2\pi\hbar(E_{0+}-E_F)\Gamma/(v^2eB)}\sim1$.

{\color{blue}\emph{Oscillation due to the inversion of the lowest Landau band beyond the QL in 3D}} --
The above discussion in 2D system can be generalized to 3D case. In a $z$-direction magnetic field $B$, we use a ${\bf k}\cdot{\bf p}$ Hamiltonian \cite{ChenRY15prl}
\begin{eqnarray}\label{Ham}
  H&=&v[(k_x+\frac{yeB}{\hbar})\tau_x\otimes\sigma_z+k_y\tau_y\otimes\mathbbm{1}_2+k_z\tau_x\otimes\sigma_x]\nonumber\\
  &&+m\tau_z\otimes\mathbbm{1}_2+\delta\mathbbm{1}_2\otimes\sigma_z,
\end{eqnarray}
where $v$ and $m$ are the Fermi velocity and mass, respectively, and $\delta= g\mu_B B/2$ is the Zeeman field. This model can describe the low-energy spectrum for a variety of Dirac materials, from semimetals to insulators. The energy spectrum is found as
\begin{align}
  E_{\nu \lambda s}=\lambda\sqrt{\left(\sqrt{\nu\omega^2+m^2}+s \delta\right)^2+v^2k_z^2},
\end{align}
for $\nu\ge1$; and
\begin{align}
  E_{0\lambda}=\lambda\sqrt{(m-\delta)^2+v^2k_z^2},
\end{align}
for $\nu=0$.
Here $\omega=v\sqrt{2 eB /\hbar}$.
Instead of the Landau levels, we have the bands of Landau levels that disperse with $k_z$, because the $k_z$ along the direction of magnetic field is a good quantum number. Each time $\delta=\sqrt{\nu\omega^2+m^2}$, $E_{\nu + -}$ crosses with $E_{\nu - -}$.
Without loss of generality, we assume the carrier is electron, the Fermi energy $E_F$ can be determined from the 3D electron density for fixed density
\begin{align}
N_e=\frac{eB}{2\pi\hbar}\sum_{k_z}\left[n_{\rm F}(E_{0+})+\sum_{\nu,s}n_{\rm F}(E_{\nu+s})\right].
\end{align}

\begin{figure}[htbp]
\centering
\includegraphics[width=0.85\columnwidth]{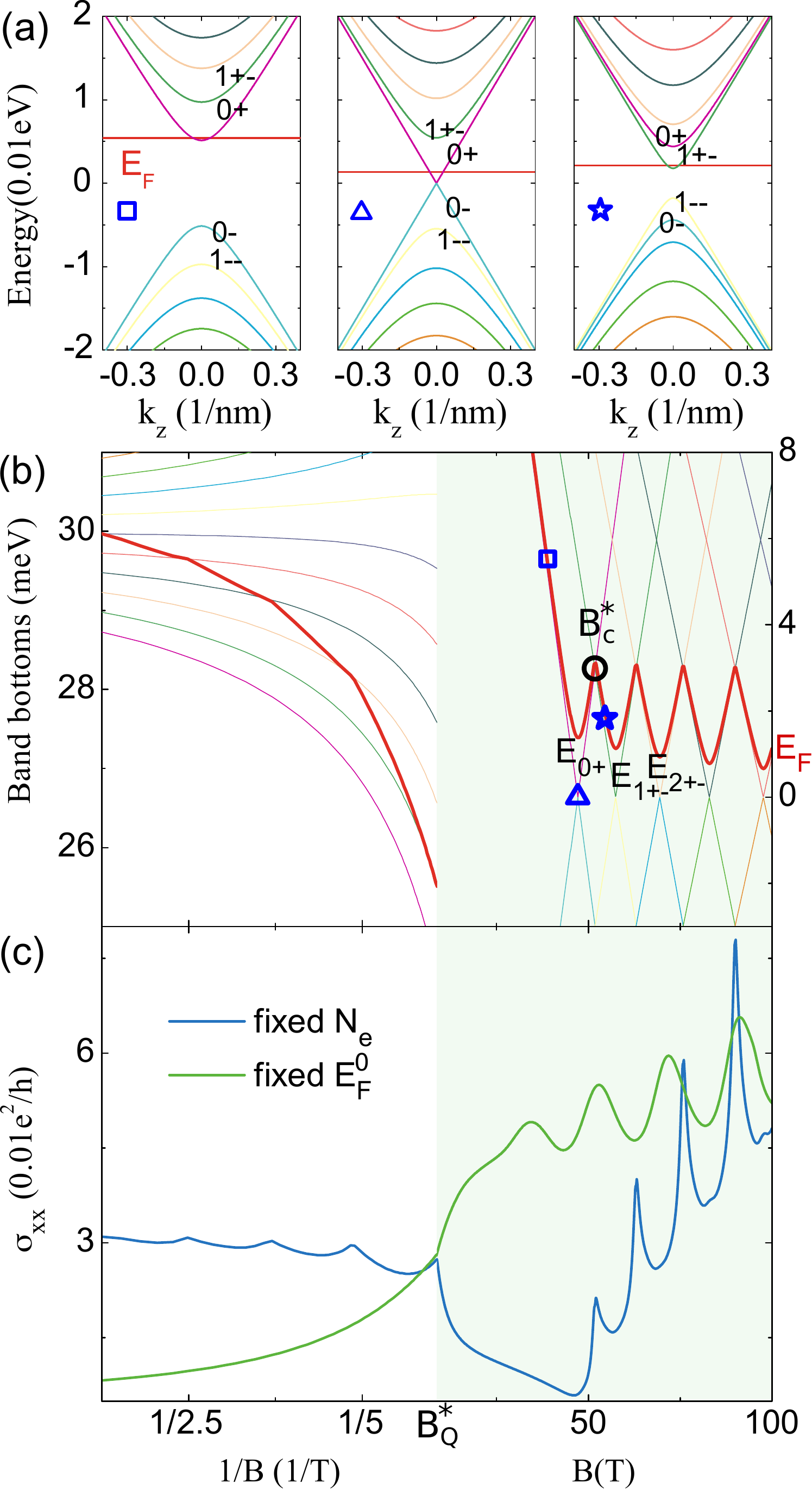}
\caption{(Color online) The quantum oscillations beyond the QL (green shadow) induced by the inversion of the lowest Landau band for the 3D Dirac model in Eq. (\ref{Ham}). (a) The 1D bands (disperse with $k_z$) of the Landau levels at magnetic fields marked by $\square$, $\triangle$, and $\medstar$ in (b).  (b) The band bottoms of  $E_{0+}$, $E_{1+-}$, $\cdots$ and the Fermi energy $E_F$ (red solid) as functions of $B$ for a fixed density $N_e$. $\bigcirc$ marks the critical field $B_c^*$ where $E_F$ is moving from $E_{0+}$ to $E_{1+-}$. (c) The conductivity $\sigma_{xx}$ as a function of the magnetic field for a fixed $N_e$ or a fixed zero-field Fermi energy $E_F^0$. $\sigma_{xx}$ for a fixed $E_{F}^0$ has been divided by 7 for display. The QL critical magnetic field  $B_{\rm Q}^*$ is 8.8 T. The SdHO is on the left with $1/B$ as the horizontal axis. The parameters are $v=0.5$ eV\AA, $m=0.03$ eV, $g=22$ \cite{ChenRY15prl,weng2014prx,liu2016nc}, $N_e=1\times10^{17}$ cm$^{-3}$, and $\Gamma=0.001$ eV. }\label{Fig:sigma}
\end{figure}

When the Zeeman splitting $\delta$ overwhelms the Landau energies $\sqrt{\nu\omega^2+m^2}$, the inversion of the lowest Landau band arises, leading to the oscillation of Fermi energy and conductivity [Fig. \ref{Fig:sigma} (b) and (c)] beyond the QL. The critical magnetic field of the inversion can be found as
\begin{eqnarray}\label{Eq:Bc-3D}
B_c^* &=& \frac{e}{\hbar} \frac{2v^2}{g^2\mu_B^2}+\frac{2m}{g \mu_B}.
\end{eqnarray}
At the magnetic field where the lowest Landau band inverts, the Fermi energy for fixed electron density indicates a dip, as shown in Fig. \ref{Fig:sigma} (b).
The amplitude of the oscillation beyond the QL is also much larger than that of the usual SdHO. For quantities sensitive to the Fermi energy, such as the thermal power, this oscillation is even more remarkable. For a fixed Fermi energy, there is also a oscillation of conductivity with larger amplitude comparing to the SdHO (the green line in Fig. \ref{Fig:sigma} (b)), but the amplitude and peak positions are quite different from those with a fixed $N_e$. 
The 3D oscillation beyond the QL is also periodic in $B$. With a large $g$ or small $v$, the spacing between the oscillation peaks can be found as
$B_T\simeq 4B_c$ just as the Fermi energy of fixed $N_e$ for small gap $m$.

%
%
%

{\color{blue}\emph{Experimental implications and discussions}} --
Distinct from other unexpected fine structures \cite{ZhangCL17np, Ramshaw18nc,Kim2017prl,wang2018sa,ChenYY18prl,Assaf17prl,Zhang2019prl,Kozii2019prb,zhang2020nc,han2019arXiv,Moll16nc, ZhangCL19nc} at extremely strong magnetic fields, our work proposes a new possible phenomenon. This $B$-oscillation is observable in Dirac materials where momentum is coupled only to pseudospin, if the magnetic field is large enough.
It can be observed at an accessible magnetic field in materials with a small Fermi velocity $v$ or large $g$ factor. To observe it at a magnetic field of $40$ T, the ratio $g/v $ has to be about 200 /eV nm.
In artificial systems that can be described by the Dirac models, such as the InAs/GaSb asymmetric quantum well and GaSb/InAs/GaSb symmetric quantum well, the electrically-tuned Fermi velocity \cite{Krishtopenko18sa,Campos2019jpcm} varies from 0.07 to 0.3 eV$\text{\AA}$ \cite{Knez2014prl,Li2015prl} and the $g$ factor is about 10 \cite{Mu2016apl}, so the ratio $g/v\in$ [330, 1400] eV$^{-1}$nm$^{-1}$, suggesting that these systems are better candidates for the oscillation beyond the QL. Moreover, the critical magnetic field for the novel inversion can be reduced by doping magnetic impurities, which effectively enhances the Zeeman $g$ factor. Also, considering the continuous growth of the largest magnetic field generated in laboratories \cite{Nakamura18rsi,Nakamura20prb}, this novel oscillation may be observed in a near future.

Here both two Dirac models contain only one Fermi pocket. For a system having $N$ pseudospin Dirac pockets with relatively small Fermi velocities $v_i$ and large $g$ factors $g_i$ ($i=1,\cdots,N$), if the distances among these pockets are small enough, the magnetic breakdown \cite{Cohen1961prl,Priestley1963pr,OBrien2016prl,Alexandradinata2017prl} due to the tunneling between pockets may occur. Beyond the QL of all pockets, this novel $B$-type oscillation will have a period of $B_T=(4e/\hbar)[\sum_i(g_i^2\mu_B^2/v_i^2)]^{-1}$ in 2D and $B_T=(8e/\hbar)[\sum_i(g_i^2\mu_B^2/v_i^2)]^{-1}$ in 3D. If only a portion of pockets are beyond the QL, the oscillation will be between $B$-type and $1/B$-type. 

This $B$-oscillation can also happen in the ballistic regime, where the conductance has nothing to do with disorder but depends on the density of states \cite{Datta2005}, so can directly probe the ILLL. Here we only consider the oscillation of the conductivity, other physical quantities such as magnetic susceptibility and magnetization in de Haas-van Alphen effect can also exhibit this novel quantum oscillation, up to some phase shifts. It should be noted that the mechanism of the ILLL is completely different from the Zeeman splitting in usual systems \cite{Cao15nc,liu2016nc,Hu2017prb,Wang2018pnas,Bi2018njp}. Although in usual systems the Landau level inversion may also happen for higher index levels with large  Zeeman splitting, the lowest one is always the zeroth level. Therefore, this oscillation does not occur in traditional semiconductors with parabolic dispersion and real-spin Dirac materials. In some cases of pseudospin Dirac materials, the magnetic field may lift negative energy Landau levels up to the Fermi energy to give rise to oscillations (Fig. 1 (a)). In other cases, the oscillation beyond the QL could have nothing to do with the negative-energy Landau levels (Fig. 2 (a)), solely as a result of the competition between orbital pseudospin and Zeeman real-spin degrees of freedom of one type of carriers. This novel phenomenon may also exist widely in topological materials with both real spin and pseudospin degrees of freedom, where the energy spectrum linearly depends on the momentum near the contacting points between conduction and valence bands.

We discuss some physical effects that may spoil this oscillation. The fractional quantum Hall effect may happen when the last Landau level is occupied in conventional two-dimensional electron gases, because of strong Coulomb interactions. For the present case, the electron-electron interaction could also be significant. However, the last level and fractional filling is not well defined as a result of the level inversion. The interaction may influence the oscillation in other ways. Other physical phenomena involving spin degree of freedom may also affect this oscillation. If the energy scales relating to these effects are smaller than the Zeeman splitting, the profile of the oscillation is supposed to be observable. The unavoidable level repulsion in real systems may open a gap near the level crossing. However, the ILLL still exists for small electron density. Therefore, the occupied lowest level still changes with the change of the field, then the $B$-oscillation of the Fermi energy or the conductivity is also expected to be observed.


{\color{blue}\emph{Acknowledgments}} --
This work was supported by the National Natural Science Foundation of China (11534001, 11974249, and 11925402), the Strategic Priority Research Program of Chinese Academy of Sciences (Grant No. XDB28000000), Guangdong Innovative and Entrepreneurial Research Team Program (2016ZT06D348), the National Key R \& D Program (2016YFA0301700), the Natural Science Foundation of Shanghai (Grant No. 19ZR1437300), Shenzhen High-level Special Fund (No. G02206304, G02206404), and the Science, Technology and Innovation Commission of Shenzhen Municipality (ZDSYS20170303165926217, JCYJ20170412152620376, KYTDPT20181011104202253). The numerical calculations were supported by Center for Computational Science and Engineering of Southern University of Science and Technology.



%

\end{document}